\RequirePackage{fix-cm}
\documentclass{svjour3}                     

\smartqed
\usepackage{graphicx,amsmath,subfig,epsfig,color,lineno,latexsym,courier}

\journalname{Transport in Porous Media}

\begin{document}
\title{A Monte Carlo Algorithm for Immiscible Two-Phase Flow in Porous Media}

\titlerunning{A Monte Carlo Method}  
\author{Isha Savani 
           \and
        Santanu Sinha 
           \and
        Alex Hansen
           \and
        Dick Bedeaux
           \and
        Signe Kjelstrup
           \and
        Morten Vassvik}

\institute{Isha Savani and Morten Vassvik \at
              Department of Physics,
              Norwegian University of Science and Technology, NTNU\\
              N-7491 Trondheim, Norway\\
              \email{Isha.Savani@gmail.com, Morten.Vassvik@ntnu.no}\\
           \and
           Santanu Sinha and Alex Hansen \at
              Beijing Computational Science Research Center\\
              10 West Dongbeiwang Road, Haidan, Beijing 100193, China\\
              \email{santanu@csrc.ac.cn, Alex.Hansen@ntnu.no}\\
           \and
           Dick Bedeaux and Signe Kjelstrup \at
              Department of Chemistry,
              Norwegian University of Science and Technology, NTNU\\
              N-7491 Trondheim, Norway\\
              \email{Dick.Bedeaux@chem.ntnu.no, Signe.Kjelstrup@ntnu.no}}

\date{Received: date / Accepted: date}

\maketitle

\begin{abstract}
We present a Markov Chain 
Monte Carlo algorithm based on the Metropolis algorithm
for simulation of the flow of two immiscible fluids in a porous medium
under macroscopic steady-state conditions using a dynamical pore
network model that tracks the motion of the fluid interfaces.  The
Monte Carlo algorithm is based on the configuration probability, 
where a configuration is defined by the positions of all fluid
interfaces. We show that the configuration probability is
proportional to the inverse of the flow rate. Using a two-dimensional
network, advancing the interfaces using time integration the computational
time scales as the
linear system size to the fourth power, whereas the Monte Carlo computational
time scales as the linear size to the second power. We discuss the
strengths and the weaknesses of the algorithm.

\keywords{dynamical pore network models, Markov Chain Monte Carlo,
Metropolis Monte Carlo, Immiscible Two-Phase Flow, Ergodicity.}
\end{abstract}

\section{Introduction}
\label{sec:intro}

The characterization of porous media at the pore level is undergoing a
revolution \cite{bbdgimpp13}. Through the use of new scanning
techniques, we are capable of reconstructing the pore space
completely, including the tracking of motion of immiscible fluids. A
gap is now appearing between the geometrical characterization of
porous media and our ability to predict their flow properties based on
this knowledge.

The pore scale may be of the order of microns whereas the
largest scales --- e.g.\ the reservoir scale --- may be measured in
kilometers. Hence, there are some eight orders of magnitude between
the smallest and the largest scales. At some intermediate scale, that
of the representative elementary volume (REV), the porous medium may
be regarded as a continuum and the equations governing the flow
properties are differential equations. The crucial problem is to
construct these effective differential equations from the physics at
the pore scale. This is the upscaling problem. A possible path towards
this goal is to use brute computational power to link the pore scale
physics to pore networks large enough so that a continuum description
makes sense. Alas, this is still beyond what can be done
numerically. However, computational hardware and algorithms are
steadily being improved and we are moving towards this goal.

It is the aim of this paper to introduce a new algorithm that improves
significantly on the efficiency of network models \cite{jh12}. These
are models that are based on the skeletonization of the spaces in such
a way that a network of links and nodes emerge. Each link and node are
associated with parameters that reflect the geometry of the pore space
they represent. The fluids are then advanced by time stepping some
simplified version of equations of motion of fluid. The bottle neck in
this approach is the necessity to solve the Kirchhoff equations to
determine the pressure field whose gradients drive the fluids in
competitions with the capillary forces.

A different and at present popular computational approach, among several, 
is the lattice Boltzmann method \cite{rob10,rino12}. 
This method, based on simultaneously solving the
Boltzmann equations for different species of lattice gases, is very
efficient compared to the network approach necessitating solving the
Kirchhoff equations. However, the drawback of the lattice Boltzmann
approach is that one needs to resolve the pore space. Hence, one needs
to use a grid with a finer mask than the network used in the network
approach. This makes the lattice Boltzmann approach very efficient at
the scale where the actual shape of the pores matter, but not at the
larger scale where the large scale topology of the pore network is
more important. Further methods which resolve the flow at the pore
level are e.g.\ smoothed particle hydrodynamics \cite{tm05,op10,ll10}
and density functional hydrodynamics \cite{abdekks16}.  When network
models are so heavy numerically that the networks that can be studied
are not much larger than those studied with the pore scale methods,
the latter win as they can give a more detailed description of the
flow.  However, if the computational limitations inherent to network
models could be overcome, they would form an important tool in
resolving the scale-up problem: at small scale network models would be
calibrated against the methods that are capable of resolving the flow
at the pore level. On large scales, their results may be extrapolated
to scales large enough for homogenization, i.e., replacing the
original pore network by a continuum.

As pointed out above, the bottleneck in the network models is the
necessity to determine the pressure field at each time step. When the
time steps are determined by the motion of the fluid interfaces, these
will be small as they typically are set by the time lapse before the
next interface reaches a node in the network. Time stepping allows
detailed questions concerning how flow patterns develop in time to be
answered. That is, the time stepping provides a detailed sequence of
configurations where each member of the sequence is the child of the
one before and the parent of the one after. If the quantities that are
calculated are averages over configurations, time stepping will
provide too much information; for averages the {\it order\/} in which
the configurations occur is of no consequence. If the order in which
the fluid configurations occur is scrambled, the averages remain
unchanged. This is where the Monte Carlo method enters. It provides a
way to produce configurations that will result in the same averages as
those obtained through time stepping. The order in which the
configurations occur will be different from those obtained by time
stepping. The time stepping procedure necessitates that there are tiny
differences between each configuration in the sequence, since the time
steps have to be small. This limitation is overcome in the Monte Carlo
method which we will describe here. This makes the Monte Carlo method
much more efficient than time stepping as we will see.

In Section \ref{sec:network} we describe the network model we use to
compare the Monte Carlo method with time stepping, see Aker et
al.\ and Knudsen et al.\ \cite{amhb98,kah02}. In the next Section
\ref{sec:metropolis}, we start by explaining the statistical mechanics
approach to immiscible two-phase flow in porous media that lies behind
the Monte Carlo algorithm we propose \cite{hsbksv16,sbksvh16}. In
particular, we derive the configuration probability --- the
probability that a given distribution of fluid interfaces in the model
will appear.  This is also known as the {\it ensemble distribution\/} 
in the statistical physics community.  Based on this knowledge, 
we then go on to describe the
Monte Carlo algorithm itself. This section is followed by Section
\ref{sec:results} where we compare the Monte Carlo method with time
stepping using the same network model described in Section
\ref{sec:network}. We then go on to compare the efficiency in terms of
computational cost of the two methods. We end this section by
discussing the limitations of the Monte Carlo algorithm as it now
stands and point towards how these may be overcome. We end by Section
\ref{sec:conc} where we summarize the work and draw our conclusions.

\section{Network Model}
\label{sec:network}

\begin{figure}
\includegraphics[width = 0.6\textwidth,clip]{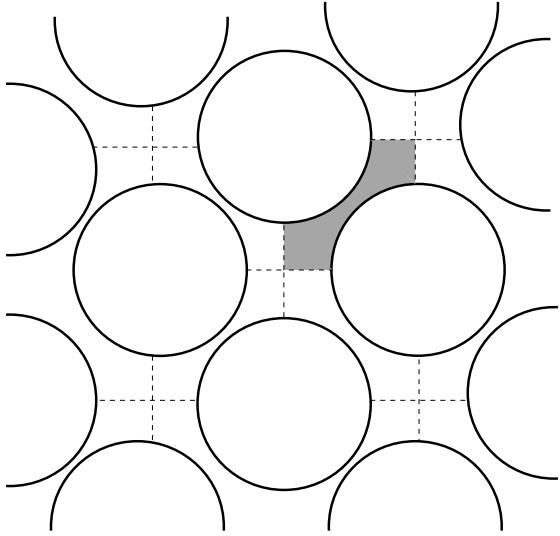}
\caption{\label{fig2-1} The geometry of the pore network we use. The
  shaded area constitutes a link between two nodes.}
\end{figure}

In order to have a concrete system to work with, we describe here the
details of the network model we use. The model is essentially the one
first developed in references \cite{amhb98,kah02}. For simplicity we
do not consider a reconstructed pore network based on a real porous
medium \cite{toh12,rho09}. Rather, we simply use a two-dimensional
square network, with disorder in the pore radii, oriented at
45$^\circ$ with respect to the average flow direction as shown in
Figure \ref{fig2-1}. As described in \cite{kah02}, we use {\it
  bi-periodic boundary conditions.\/} Hence, the network takes a form
of the surface of a torus. In this way, the two-phase flow enters a
steady state after an initial transient period. This steady state does
{\it not\/} mean that the fluid interfaces are static. Rather, we use
capillary numbers high enough so that fluid clusters incessantly form
and break up. By {\it steady state\/} we mean that the macroscopic
averages --- averages over the entire network --- are well defined and
do not drift.

The network contains $L\times L$ links. All links have equal length
$l$, but their radii have been drawn from a uniform distribution of
random numbers in the interval $[0.1l,0.4l]$. We set $l=1$mm. We
neglect gravitational effects.

\begin{figure}
\includegraphics[width = 0.6\textwidth,clip]{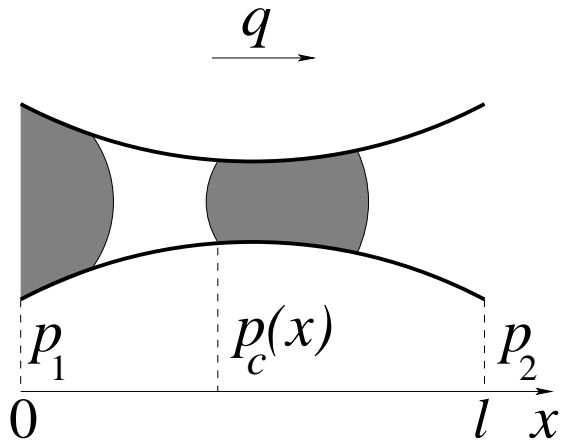}
\caption{\label{fig2-2} This is one of the links in the network. The
  wetting and non-wetting fluids, coloured by white and gray
  respectively, are seperated by interfaces. Each interface provides a
  capillary pressure $p_c(x)$ that point in the direction from the
  non-wetting towards the wetting fluid. Through the link a flow $q$
  passes. We indicate the two node pressures $p_1$ and $p_2$ at the
  end of the link.}
\end{figure}

Fluid flow through each link in the network is modeled using the
Washburn equation \cite{w21}, see Figure \ref{fig2-2}. There is a
volume flow $q$ passing through it driven by the two pressures $p_1$
and $p_2$. Each fluid interface contributes a capillary pressure
$p_c(x)$ where $x \in [0,l]$ is the position of the interface. The
capillary pressure is given by the Young-Laplace equation
\begin{equation}
\label{eq:cap}
|p_c(x)| = \frac{2\gamma \cos \theta}{r_0}
\left[1-\cos \left(2\pi \frac{x}{l}\right)\right]\;,
\end{equation}
where $\gamma$ is the surface tension, $\theta$ the contact angle
between the interface and the pore wall. We set $\gamma\cos\theta=30$
dyn/cm. $r_0$ is the average link radius. We assume that the link has
a shape so that $p_c$ attains the given $x$ dependence. It has been
chosen so that $p_c(0)=p_c(l)=0$ and $\max_x |p_c(x)|=|p_c(l/2)|$. The
Washburn equation then becomes
\begin{equation}
\label{eq:wash}
q=-\frac{\pi r_0^{4}}{8\mu _{av}}\left[p_2-p_1- \sum_i p_{c}(x_i)\right]\;,  
\end{equation}
where $\mu_{av}=s_{nw}\mu_{nw}+s_w\mu_w$ is the
viscosity. $s_{nw}=l_{nw}/l$ and $s_{w}=l_w/l$ are the fractions of
the link length that cover the non-wetting and wetting fluids
respectively so that $s_{nw}+s_w=1$. We set $\mu_{nw}= \mu_w=1$ poise.

We define the capillary number $\rm Ca$ as 
\begin{equation}
\label{eq:ca}
{\rm Ca} = 
\frac{\langle |q|\rangle \langle \mu_{av}\rangle}
{\gamma \pi \langle r_0\rangle^2}\;,
\end{equation}
where $\langle \cdots\rangle$ is an average over all links.

A pressure difference $\Delta P$ is applied across the network. This
is done in spite of the network being periodic in the direction of the
pressure difference, see Knudsen et al.\ \cite{kah02}. By demanding
balance of flow at each node using the Washburn equation
(\ref{eq:wash}), we determine the pressures ($p_i$) at the nodes. This
is done by solving the corresponding matrix inversion problem by using
the conjugate gradient algorithm \cite{bh88}.

When the pressures at nodes are known, the flow $q_{ij}$ --- here
between neighboring nodes $i$ and $j$ connected by a link --- is
calculated using equation (\ref{eq:wash}). Knowing the velocity of the
interfaces in each link, we then determine the time step such that any
meniscus can move a maximum distance, say, one-tenth of the length of
corresponding link in that time. All the interfaces are then moved
accordingly and the pressure at the nodes are determined again by
conjugate gradient algorithm. This is equivalent to event-driven
molecular dynamics. When an interface reaches the node, the interface
will spread into the links that are connected to the node and which
have fluid entering them from the node. The rules for how this is done
are described in detail in Knudsen et al. \cite{kah02}.

\section{Metropolis Monte Carlo}
\label{sec:metropolis}

We first describe the theory that lies behind the Monte Carlo
algorithm that we present. We need to introduce the concepts of {\it
configuration}, and {\it configuration probability,} also known as
the ensemble distribution in the statistical mechanics community. 
We then go on to derive the configurational probability. 
Armed with this, we construct the Monte Carlo
algorithm \cite{ptvf07} after having presented a short 
review of the Metropolis version of Monte Carlo \cite{ptvf07,mrrtt53}.

\subsection{Statistical Mechanics of Immiscible Two-Phase Flow}
\label{sub:statmech}

Sinha et al.\ \cite{shbk13} studied the motion of bubbles in a single
capillary tube with varying radius. Suppose that the capillary tube
has a length $L$ and a radius that varies as $r=r_0/[1-a\cos 2\pi
  x/l]$ where $l\ll L$, $a$ is an amplitude and $r_0$ the average
radius of the tube. Suppose furthermore that the tube is filled with
wetting fluid except for a bubble of length $\Delta x_b$ and a center
position $x_b$. By using equation (\ref{eq:cap}), one derives the net
capillary force from the two interfaces that limit the bubble as,
\begin{equation}
\label{eq:capbub}
p_b(x_b)=-\sigma\ \sin\left(\frac{2\pi x_b}{l}\right)\;,
\end{equation}
where $\sigma=4a\gamma\cos\theta\sin(\pi\Delta x_b/l)/r_0$. By
combining this equation with the Washburn equation (\ref{eq:wash}),
one finds
\begin{equation}
\label{eq:mot1d}
\dot{x}_b=-\frac{r_0}{8l\mu_{av}}\ 
\left[\Delta p + \sigma\sin\left(\frac{2\pi x_b}{l}\right)\right]\;,
\end{equation}
where $\pi r_0^2\dot{x}_b=q$, and $\Delta p=(L/l)\Delta P$ where $\Delta P$
is the pressure difference across the tube. 

Suppose there is a quantity $f=f(x_b)$ that depends on the position of
the bubble in the capillary tube. For example, $f$ might be the flow
$q$. Let us now assume that $\Delta P$ does not vary in time. The time
average of $f$ is then
\begin{equation}
\label{eq:timeave}
\overline{f}=\frac{1}{T_b}\ \int_0^{T_b}f(x_b(t))\ dt\;,
\end{equation}
where $x_b(t)$ is the time integration of the Washburn equation
(\ref{eq:wash}) and the time period $T_b=(2\pi\sigma)/\sqrt{\Delta
  p-\sigma^2}$. We note, {\it and this is the crucial observation,\/} that we
may change integration variable from time $t$ to bubble position
$x_b$,
\begin{equation}
\label{eq:fxave}
\overline{f}=\frac{1}{T_b}\ \int_0^l f(x_b)\ \frac{dx_b}{dx_b/dt}
=\int_0^l f(x_b)\Pi(x_b)dx_b\;,
\end{equation}
where
\begin{equation}
\label{eq:1dpi}
\Pi(x_b)=\frac{1}{T_b(dx_b/dt)}=\frac{\pi r_0^2}{T_b}\ \frac{1}{q}
\end{equation}
is the {\it configuration probability.\/} That is, the {\it
  configuration\/} of the tube is given by the position of $x_b$ of
the bubble. Equation (\ref{eq:1dpi}) gives the probability density to
find the bubble at position $x_b$ --- and hence in that configuration.

The Washburn equation (\ref{eq:mot1d}) gives the motion of the bubble 
that is used in equations (\ref{eq:fxave}) and (\ref{eq:1dpi}).  The
Washburn equation assumes that we control the pressure drop $\Delta P$.  
If we on the other hand control the flow $q$, the equation of motion
becomes
\begin{equation}
\label{eq:mot1dq}
\dot{x}_b=\frac{q}{\pi r_0^2}\;.
\end{equation}
The time period now becomes
\begin{equation}
\label{eq:timeperiodq}
T_b=\frac{\pi r_0^2 L}{q}\;,
\end{equation}
and hence the configurational probability is
\begin{equation}
\label{eq:1dqi}
\Pi(x_b)=\frac{\pi r_0^2}{T_b}\ \frac{1}{q}=\frac{1}{L}\;,
\end{equation}
which states that all positions of the bubble is equally probable.  

To ramp up the complexity of the problem, we assume that there are
$N$ bubbles in the one-dimensional tube. The centers
of mass of bubble number $j\in[1,N]$ is $x_j$ and it has a width of 
$\Delta x_j$.  Since the system is one dimensional, all bubbles move with 
the same speed $\dot{x}_j=\dot{x}_1$.  The Washburn equation is then
\begin{equation}
\label{eq:washmany}
\dot{x}_j=\dot{x}_1=-\frac{r_0}{8L\mu_{av}}\ 
\left[\Delta p + \sum_{j=1}^N
\gamma_j\sin\left(\frac{2\pi}{l}(x_1+\delta x_j)\right)\right]\;,
\end{equation}
where $\delta x_j=x_j-x_1$ and
\begin{equation}
\label{eq:gammaj}
\gamma_j=\frac{4\sigma a}{r_0}\ \sin\left(\frac{\pi \Delta x_j}{l}\right)\;.
\end{equation}
Solving the equations of
motion (\ref{eq:washmany}) gives $x_j=x_j(t)$.  We may invert $x_1=x_1(t)$
to get $t=t(x_1)$.  Hence, we then have $x_j(x_1)=x_j(t(x_1))$ for all $j$.
Suppose now we have a function $f=f(x_1,\cdots, x_N)$, analogous to the
one introduced in equation (\ref{eq:fxave}).  Its time average is
\begin{eqnarray}
\label{eq:fxavemany}
\overline{f}&=&\frac{1}{T_b}\int_0^{T_b} 
f\left(x_1(t),\cdots, x_N(t)\right)\ dt\nonumber\\
&=&\frac{1}{T_b}\ \int_0^L 
f\left(x_1,\cdots, x_N(x_1)\right)
\ \frac{dx_1}{dx_1/dt}\nonumber\\
&=&\int_0^L 
f\left(x_1,\cdots, x_N(x_1)\right)
\Pi(x_1)dx_1\;,\nonumber\\ 
\end{eqnarray}
where
\begin{equation}
\label{eq:1dpimany}
\Pi(x_1)=\frac{1}{T_b}\frac{1}{(dx_1/dt)}=\frac{\pi r_0^2}{T_b}\ \frac{1}{q}\;,
\end{equation}
where $q=\pi r_0^2 \dot{x}_1$. This is precisely the same expression as in
(\ref{eq:1dpi}).

We now turn to complex network topologies.  For concreteness, we may
imagine a two-dimensional square network.  However, the arguments
presented in the following are general.  A configuration is given
by the position of all interfaces. Let us denote that $\vec
x=(x_1,x_1,x_2,\cdots,x_N)$, where $x_i$ is the position of the $i$th
interface. Hence, $x_i$ contains information both on which link the interface
sits in and where it sits in the link.  
A flow $Q$ passes through the network. The flow equations for the
network consist of a Washburn constitutive equation for each link combined
with the Kirchhoff equations distributing the flow between the links. The
motion of the interfaces are highly non-linear, but 
of the form $\dot{x}_i=g_i(\vec x)$.  Solving these equations
gives $x_j=x_j(t)$. 

Again we consider a function $f=f(\vec x)$ of the position of the interfaces.
Its time average is 
\begin{eqnarray}
\label{eq:fxcomplex}
\overline{f}&=&\frac{1}{T_b}\int_0^{T_b} 
f\left(\vec x(t)\right)\ dt
=\frac{1}{T_b}\ \int_0^L 
f\left(\vec x(x_i)\right)\ \frac{dx_i}{dx_i/dt}\nonumber\\
&=&\int_0^L 
f\left(\vec x(x_i)\right)\ \Pi(x_i) dx_i\;.\nonumber\\
\end{eqnarray}
Here we have inverted $x_i=x_i(t)$ so that we have $t=t(x_i)$ 
and then substituted $\vec x(t)=\vec x(t(x_i))=\vec x(x_i)$.
The configurational probability is defined as before,
\begin{equation}
\label{eq:dpicomplex}
\Pi(\vec x)=\frac{1}{T_b}\ \frac{1}{dx_i/dt}\;.
\end{equation}
Let us now choose $x_i=x_1$ to be an interface moving in a link that carries
all the flow in the network.  Such a link is a capillary tube 
connected in series with the rest of the network.  In this case we
have $\dot{x_1}=Q/\pi r_0^2$, where $Q$ is the total flow.  Hence,
we have 
\begin{equation}
\label{eq:piintq0}
\Pi(x_1)=\frac{\pi r_0^2}{T_b}\ \frac{1}{Q}\;.
\end{equation}

We have in the discussion so far compared the time evolution of a given
sample defined by an initial configuration of interfaces.   We now 
imagine {\it an ensemble\/} of initial configurations of interfaces. 
Each sample evolves in time and there will be a configurational probability
(\ref{eq:piintq0}) for each.  This will have the same value 
for each configuration
$\vec x$ that corresponds to the same flow $Q$. Hence, we have the 
configurational probability
\begin{equation}
\label{eq:piintq}
\Pi(\vec x) \propto \frac{1}{Q}\;.
\end{equation}
This equation is the major theoretical result of this paper: all
configurations corresponding to the same $Q$ are equally probable.
Intuitively, equation (\ref{eq:piintq}) makes sense: The slower the
flow, proportionally the more the system stays in --- or close to ---
a given configuration \cite{sbkvhs16}.

Is the system ergodic? Equations (\ref{eq:fxave}), (\ref{eq:fxavemany})
and (\ref{eq:fxcomplex}) answer this question
positively. Time averages give, by construction, the same results as
configurational averages.

\subsection{Implementation of the Metropolis Algorithm}
\label{sub:implementation}

In order to present the details of the Metropolis Monte Carlo
algorithm that we propose, we first review the general formulation of
the Metropolis algorithm \cite{k06,lb15}.

\subsubsection{General Considerations}
\label{subsub:general}

We have a set of configurations characterized by the variable $\vec
x$, the positions of the interfaces. We now wish to construct a {\it
  biased random walk\/} through these configurations so that the
number of times each configuration is visited --- i.e., the random
walk comes within $d{\vec x}$ of the configuration --- is proportional
to $\Pi({\vec x})$. proportional to the probability for that
configuration. The Metropolis algorithm accomplishes this goal. In
order to do so, a transitional probability density from state $\vec x$
to state $\vec x'$ is constructed as
\begin{equation}
\label{eq:met}
\Pi(\vec x,\vec x')=\pi(\vec x,\vec x')\ 
\min\left(1,\frac{\Pi(\vec x')}{\Pi(\vec x)}\right)\;.
\end{equation}
where $\pi(\vec x,\vec x')$ is the probability density to pick trial
configuration $\vec x'$ given that the system is in configuration
$\vec x$. It is crucial that $\pi(\vec x', \vec x)$ is symmetric,
\begin{equation}
\label{eq:sym}
\pi(\vec x,\vec x')=\pi(\vec x',\vec x)\;.
\end{equation}
Equations (\ref{eq:met}) and (\ref{eq:sym}) ensure detailed balance,
\begin{equation}
\label{eq:bal}
\Pi(\vec x)\Pi(\vec x,\vec x')=\Pi(\vec x')\Pi(\vec x',\vec x)\;.
\end{equation}
Detailed balance guarantees that the biased random walk visits the 
configurations $\vec x$  with a frequency proportional to $\Pi(\vec x)$.
The generated configurations follow the ensemble distribution.

\begin{figure}
\includegraphics[width = 0.6\textwidth,clip]{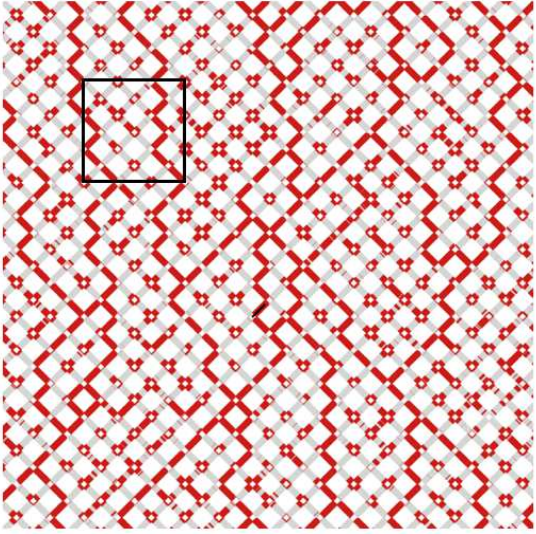}
\caption{\label{fig3-1} We show here a typical network of the kind we
  use for comparing the time stepping and Monte Carlo methods. The
  network is bi-periodic and the flow is from the bottom towards the
  top.  The dark red constitutes the non-wetting fluid and the gray
  constitutes the wetting fluid. When using the Monte Carlo method, a
  random sub network is chosen as shown in the box, taken out of the
  network, integrated forward in time after having been made
  bi-periodic, and then re-entered into the network. This is the heart
  of the Metropolis Monte Carlo algorithm.}
\end{figure}

When we combine equations (\ref{eq:piintq}) and (\ref{eq:met}),
we have 
\begin{equation}
\label{eq:metq}
\Pi(\vec x,\vec x')=\pi(\vec x,\vec x')\ 
\min\left(1,\frac{Q(\vec x)}{Q(\vec x')}\right)\;.
\end{equation}

\subsubsection{The Implementation}
\label{subsub:implementation}

The Metropolis Monte Carlo algorithm based on equation (\ref{eq:metq})
consists of two crucial steps. The first step consists in generating a
{\it trial configuration\/} and the second step consists in deciding
whether to keep the old configuration or replacing it with the trial
configuration.

The first step, generating the trial configuration, is governed by the
trial configuration probability $\pi(\vec x,\vec x')$ which must obey
the symmetry (\ref{eq:sym}). That is, if the system is in
configuration $\vec x$, the probability to pick a trial configuration
$\vec x'$ must be equal to the probability to pick as trial
configuration $\vec x$ if the system is in configuration $\vec x'$.

Suppose the system is in configuration $\vec x$. One needs to define a
{\it neighborhood\/} of configurations among which the trial
configuration is chosen. If the neighborhood is too restricted, the
Monte Carlo random walk will take steps that are too small and hence
would be inefficient. If, on the other hand, the neighborhood is too
large, the random walk ends up doing huge steps that will miss the
details.

We propose generating the trial configurations as follows. Our system
is shown in figure \ref{fig3-1} and consists of $L\times L$ links as
described in Section \ref{sec:network}. There is a flow $q_{ij}$
through link $ij$ connecting the neighboring nodes $i$ and $j$. There
is a total flow rate $Q$ in the network given by
\begin{equation}
\label{eq:Q}
Q=\frac{1}{L}\ \sum_{{\rm all}\ ij} q_{ij}\;,
\end{equation}
and a corresponding pressure drop $\Delta P$.  

We choose a randomly positioned sub network as shown in figure
\ref{fig3-1}. The network consists of $\Lambda\times\Lambda$ links. We
``lift" the sub network out of the complete network and fold it into a
torus, i.e, implementing bi-periodic boundary conditions. The
configurations of fluid interfaces in the sub network remains
unchanged at this point.

We calculate the flow rate in the sub network 
\begin{equation}
\label{eq:subQ}
\Theta=\frac{1}{\Lambda}\ \sum_{ij\ {\rm in\ sub\ network}} q_{ij}\;.
\end{equation}
By solving the Kirchhoff equations on the sub network, we {\it time
  step\/} the configuration forwards in time while keeping the flow
rate $\Theta$ constant. We end the time integration when $4$ ---
arbitrarily chosen --- sub network pore volumes have passed through
it.

The bi-periodic boundaries of the sub network is then opened up and
the sub network with the new configuration of fluid interfaces is
placed back into the full network. This is then the trial
configuration $\vec x'$.

Part of the probabilistic choice of the trial configuration that
defines $\pi(\vec x,\vec x')$ rests on the choice of the sub network:
its position is picked at random. Hence, if the system is in state
$\vec x$ or in trial state $\vec x'$, the probability to pick a
particular sub network is the same. This makes this part of the choice
of trial configuration symmetric. When the sub network is time stepped
for $4$ sub system pore volumes, this is done at constant flow rate
$\Theta$. Hence, all sub network configurations are equally probable,
see equation (\ref{eq:piintq}). Hence, also this part of the choice of
trial configuration is symmetric. The full probability $\pi(\vec
x,\vec x')$ is the probability of picking a given sub network times
the probability that a given configuration will occur. Combining the
two leads to the necessary symmetry (\ref{eq:sym}).

We point out here that whereas the configurational probability $\Pi(\vec x)$
in (\ref{eq:piintq}) is valid for all configurations, through
the way we generate our samples, we are restricting ourselves to physically
realistic samples in that they are generated through time stepping
parts of the system.  We cannot at this stage prove that this does not
bias our sampling. 

Once the trial configuration $\vec x$ has been generated, it is
necessary to calculate the total flow rate $Q=Q(\vec x')$ in the
network. We then decide to accept the trial configuration $\vec x'$
by using (\ref{eq:metq}). This defines a Monte Carlo {\it update.\/}

We repeat this procedure until each link in the network has been part
of at least one sub network. This defines a Monte Carlo {\it sweep.\/}

\section{Results}
\label{sec:results}

\begin{figure}
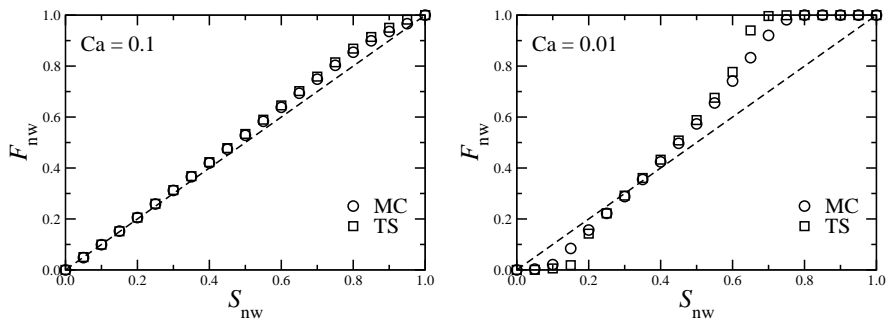

\centerline{\hfill
\includegraphics[width = 0.46\textwidth,clip]{fig4-1a.eps}\hfill
\includegraphics[width = 0.46\textwidth,clip]{fig4-1b.eps}\hfill}
\caption{\label{fig4-1} Non-wetting fractional flow ($F_\text{nw}$) as
  a function of non-wetting saturation ($S_\text{nw}$) in the steady
  state obtained via Monte Carlo simulations (MC) with constant flow
  rate ($Q$) at capillary numbers $\text{Ca}=0.1$ and $0.01$. Results
  are compared with that obtained via time stepping simulations
  (TS). The diagonal dashed lines in the plots imply
  $F_\text{nw}=S_\text{nw}$, a system of miscible fluids would follow
  that line. The data are averaged over 10 samples.}
\end{figure}

\begin{figure}
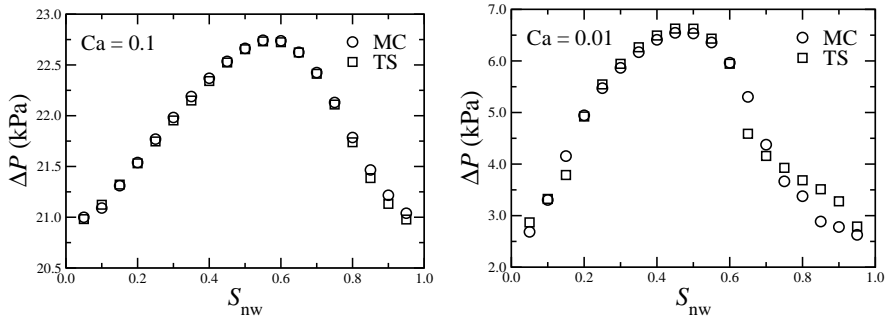

\centerline{\hfill
\includegraphics[width = 0.46\textwidth,clip]{fig4-2a.eps}\hfill
\includegraphics[width = 0.46\textwidth,clip]{fig4-2b.eps}\hfill}
\caption{\label{fig4-2} Values of pressure difference ($\Delta P$) for
  constant flow rate ($Q$) in the steady state as a function of
  non-wetting saturation ($S_\text{nw}$) for the capillary numbers
  $\text{Ca}=0.1$ and $0.01$ obtained via Monte Carlo (MC) simulations
  and time stepping (TS). The data are averaged over 10 samples.}
\end{figure}

We now present numerical results of the Monte Carlo simulation
considering the model described in Section \ref{sec:network} and we
will compare them with the results by time stepping
simulations. Simulations are performed for two different ensembles,
one is when the total flow rate $Q$ is kept constant (CQ ensemble) and
the other when the total pressure drop $\Delta P$ is kept constant (CP
ensemble). A network of $40\times 40$ links ($L=40$) is considered for
both Monte Carlo and time stepping procedure. The sub network size is
$20\times20$ links ($\Lambda=20$). To identify whether the system has
reached the steady state, we measured the quantities as a function of
time steps in time stepping and as a function of sweeps in the case of
Monte Carlo. We then identified the steady states when the averages of
measured quantities (eg. $F_\text{nw}$ and $\Delta P$ or $Q$) did not
change with time or with sweeps. We then take average over time (time
stepping) or sweeps (Monte Carlo) which give us the time average and
the ensemble average, respectively. We average 10 different networks,
but with the same sequence of networks for both Monte Carlo and time
stepping. First we present the results for CQ ensemble. Two capillary
numbers, $\text{Ca} = 0.1$ and $0.01$ are used, and for each
$\text{Ca}$, simulations are performed for different values of
non-wetting saturations in intervals of $0.05$ from 0.05 to 0.95.

\subsection{Constant $Q$ ensemble}
\label{sub:cq}

With  $Q$ constant, the Metropolis Monte Carlo algorithm becomes very 
simple. Equation (\ref{eq:metq}) simply becomes
\begin{equation}
\label{eq:metq-simple}
\Pi(\vec x,\vec x')=\pi(\vec x,\vec x')\;. 
\end{equation}
In other words, all trial configurations are accepted.

In figure \ref{fig4-1} we plot $F_\text{nw}$ --- the non-wetting
fractional flow --- as a function of $S_\text{nw}$ --- the non-wetting
saturation --- where the circles and the squares denote the results
from Monte Carlo and time stepping, respectively. The plots, as
expected, show an S-shape. This is because the two immiscible fluids
do not flow equally, and the one with higher saturation
dominates. Hence, the curve does not follow the diagonal dashed line,
which corresponds to $F_\text{nw}=S_\text{nw}$, shown in the
figure. Rather, $F_\text{nw}$ is less than $S_\text{nw}$ for low
values of $S_\text{nw}$ and higher than $S_\text{nw}$ for higher value
of $S_\text{nw}$. It therefore crosses the $F_\text{nw}=S_\text{nw}$
line at some point, which is not at $S_\text{nw}=0.5$. This is due to
the asymmetry between the two fluids, as one is more wetting than the
other with respect to the pore walls. This behaviour is more prominent
for the lower value of $\text{Ca}$, as capillary forces play a more
dominant role. The curves from the Monte Carlo and time stepping
calculations fall on top of each other for most of the lower to
intermediate range of the saturation values and we only see some
difference at very high or low $S_\text{nw}$. We will present a more
quantitative comparison between the results of Monte Carlo and time
stepping later in Section \ref{sub:limitations}. The variation of
total pressure drop $\Delta P$ for the two capillary numbers as a
function of $S_\text{nw}$ are shown in figure \ref{fig4-2}. Similar to
the fractional flow plots, we see that the results are same for Monte
Carlo and time stepping for a wide range of $S_\text{nw}$. We only see
differences at high values of $S_\text{nw}$. $\Delta P$ increases with
$S_\text{nw}$, reaching a maximum at some intermediate saturation and
then decreases again. When $S_\text{nw}$ increases from zero, more and
more interfaces appear in the system causing an increase in capillary
barriers associated with interfaces. As the total flow rate $Q$ is
constant, a higher pressure is needed to overcome the capillary
barriers. The decrease of $\Delta P$ after the maximum is due to the
decrease of the number of interfaces blocking the fluids.

\subsection{Constant $\Delta P$ ensemble}
\label{sub:cp}

We now turn to the constant pressure ensemble. Here we keep $\Delta P$
constant throughout the calculations. In this case, the Metropolis Monte
Carlo algorithm, equation (\ref{eq:metq}), becomes 
\begin{equation}
\label{eq:metq-press}
\Pi(\vec x,\vec x')=\pi(\vec x,\vec x')\ 
\min\left(1,\frac{Q(\vec x,\Delta P)}{Q(\vec x',\Delta P)}\right)\;.
\end{equation}

\begin{figure}
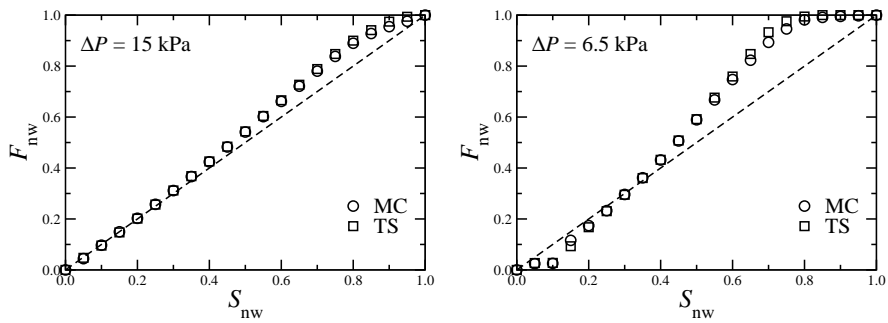

\centerline{\hfill
\includegraphics[width = 0.46\textwidth,clip]{fig4-3a.eps}\hfill
\includegraphics[width = 0.46\textwidth,clip]{fig4-3b.eps}\hfill}
\caption{\label{fig4-3} Non-wetting fractional flow ($F_\text{nw}$) as
  a function of non-wetting saturation in the steady state for
  constant $\Delta P$ ensemble. Results are presented for for Monte
  Carlo (MC) and time stepping (TS) for two different overall pressure
  drops $\Delta P = 15 \text{kPa}$ and $6.5 \text{kPa}$. As $Q$ varies
  with saturation for constant $\Delta P$, $\text{Ca}$ is not constant
  here, which is demonstrated in the next figure \ref{fig4-4}. The
  data are averaged over 10 samples.}
\end{figure}

\begin{figure}
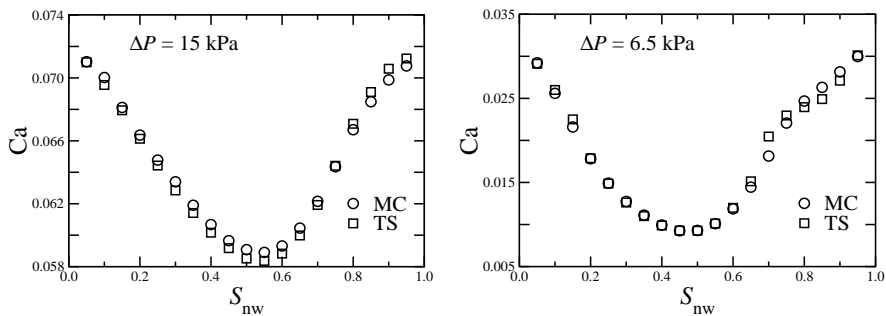

\centerline{\hfill
\includegraphics[width = 0.46\textwidth,clip]{fig4-4a.eps}\hfill
\includegraphics[width = 0.46\textwidth,clip]{fig4-4b.eps}\hfill}
\caption{\label{fig4-4} Capillary numbers, calculated from the total
  flow rates ($Q$), in the steady state as a function of the
  non-wetting saturation $S_\text{nw}$ for constant $\Delta P$
  ensemble. Results are compared between Monte Carlo (MC) and time
  stepping (TS). The data are averaged over 10 samples.}
\end{figure}

Results for the simulations with constant $\Delta P$ are shown in
figures \ref{fig4-3} and \ref{fig4-4}. Simulations are performed for
two different values of $\Delta P$, $15 \text{kPa}$ and $6.5
\text{kPa}$. The steady-state values of $F_\text{nw}$ show similar
variation with $S_\text{nw}$ as in the constant $Q$ ensemble and we
see good agreement between the results for Monte Carlo and time
stepping for a wide range of $S_\text{nw}$. Here $Q$ varies with the
saturation and the corresponding capillary numbers are plotted in
figure \ref{fig4-4} for Monte Carlo and time stepping. As discussed
before, the number of interfaces first increase with the increase in
saturation from zero, reaches a maximum value, and then decreases
again as $S_\text{nw}$ approaches $1$. The pressure is constant here,
so the total flow rate decreases with increasing capillary barriers at
the interfaces and correspondingly $\text{Ca}$ varies as in figure
\ref{fig4-4}. Here again, good match between the results Monte Carlo
and time stepping can be observed.

We show in Table \ref{table1} the percentage of rejections for the
data shown in Figure \ref{fig4-4}. The number of rejections is in all
cases quite small.  This can be understood as follows.  Set 
$Q(\vec x,\Delta P)=Q$ and $Q(\vec x',\Delta P)=Q+\delta$ where $\delta$ may
be positive or negative.  Hence, the probability to accept the new 
configuration is
\begin{equation}
\label{eq:rejection}
\min\left(1,\frac{Q(\vec x,\Delta P)}{Q(\vec x',\Delta P)}\right)=
\min\left(1,1-\frac{\delta}{Q}\right)\;,
\end{equation}
where we have assumed $\delta \ll Q$. With a small $\delta$ the probability
to reject the trial configuration is small.  This is reflected in Table 
\ref{table1}.

\begin{table}
\begin{center}
\begin{tabular}{|c|c|c|}
\hline
$\Delta p$    & $S_{nw}$ & Rejections \\ \hline
              & 0.3      &  2.1\% \\
15 kPa        & 0.5      &  2.3\% \\
              & 0.7      &  1.5\% \\ \hline 
              & 0.3      &  8.8\% \\
6.5 kPa       & 0.5      & 11.6\% \\
              & 0.7      &  4.2\% \\ \hline 
\end{tabular}
\end{center}
\caption{The percentage of rejected configurations in the
constant $\Delta P$ ensemble.
\label{table1}}
\end{table}

\subsection{Computational Cost}
\label{sub:cost}

\begin{figure}[b]
\includegraphics[width = 0.66\textwidth,clip]{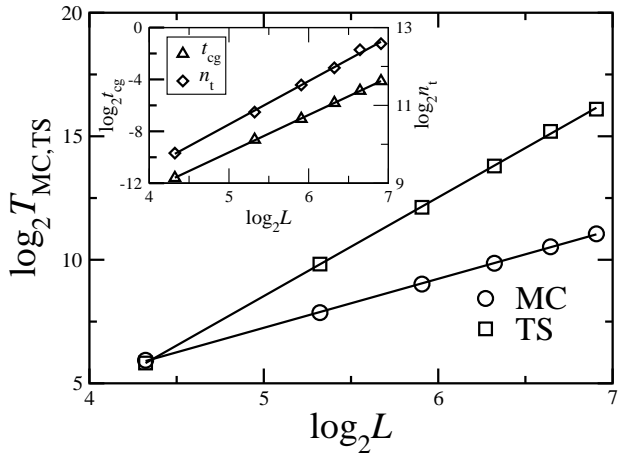}
\caption{\label{fig4-5} Plot of total computational time,
  $T_\text{MC,TS}$ (in seconds) used by Monte Carlo (MC) and time
  stepping (TS) for different system sizes ($L$). Here the time
  stepping procedure is run for $100$ injected pore volumes and in the
  Monte Carlo method, we do $25$ sweeps. Each update is based on
  running the sub system for $4$ injected sub network pore volumes. In
  this way, when $\Lambda=L=20$, the timing of the two methods are
  equal. We use the CQ ensemble with $\text{Ca}=0.1$ and
  $S_\text{nw}=0.4$. Six different system sizes, $L=20$, $40$, $60$,
  $80$ and $120$ are considered. From the slopes, the exponents
  $\alpha$ for time stepping and Monte Carlo are found. For Monte
  Carlo, we find $\alpha_\text{MC}= 1.98\pm 0.01$, which is close to
  the theoretically expected value $T_\text{MC}\sim L^2$ (see
  text). However, for time stepping, we find $\alpha_\text{TS} =
  3.99\pm 0.03$ which is much smaller than theoretical expectation --
  $T_\text{TS}\sim L^5$. In the inset, we plot the average time,
  $t_\text{cg}$, taken by the conjugate gradient solver to solve one
  entire pressure field. We find $t_\text{cg}\sim
  L^{2.88\pm0.02}$. The number of time steps per pore volume,
  $n_\text{t}$, scales as $n_\text{t}\sim L^{1.11\pm0.03}$. Combining
  these two results, we find that the computational time for the time
  stepping procedure to scales as $T_\text{TS}\sim L^{3.99}$.}
\end{figure}

Here we present a detailed comparative analysis of the computational
cost of the two algorithms. We do this by measuring the computational
time ($T_\text{MC}$ for the Monte Carlo method and $T_\text{TS}$ for
the time stepping method respectively) for different system sizes $L$.

We use the conjugate gradient method to solve the Kirchhoff equations.
This is an iterative solver. When the network contains $L\times L$
links ($L^2/2$ nodes), each iteration demands $L^2/2$ operations. The
number of iterations necessary to solve the equations {\it exactly\/}
scales as $L^2$, making the total cost scale as $L^\beta$, where
$\beta=2+2=4$.  However, in practice, the number of iterations
necessary to reach the solution of the Kirchhoff equations to within
machine precision is much lower than that needed for the theoretically
exact solution. As we shall see, $\beta$ is much smaller than four.

The number of time steps needed to push one pore volume through the
network is $n_\text{t}$. We expect it to depend on $L$ as
$n_\text{t}=aL^\tau$, where $a$ is a prefactor essentially measuring
the number of time steps on the average it takes for an interface to
cross a link. In our calculations, this is of the order of 10.
Intuitively, this number should be proportional to the width of the
network, $L$, making $\tau=1$. In practice, as we shall see, it is
slightly larger.

For each time step, the conjugate gradient demands $t_\text{cg}=b
L^\beta$ operations where $b$ is another prefactor. The total
computational time ($T_\text{TS}$) per pore volume is then
\begin{equation}
\label{timets}
T_\text{TS} = n_\text{t}\times t_\text{cg} = abL^{\tau+\beta}
=abL^{\alpha_\text{TS}}\;,
\end{equation}
where $\alpha_\text{TS}=\tau+\beta$. Based on the theoretical
considerations above, setting $\beta=4$ and $\tau=1$, we have
$T_\text{TS}\sim L^5$. The actual computational time measured using
the \texttt{clock()} function in C is plotted in figure \ref{fig4-5}
for $\text{Ca}=0.1$ and $S_\text{nw}=0.4$. We find that $T_\text{TS}$
scales with $L$ with an exponent $\alpha_\text{TS}=3.99\pm0.03$ which
is much smaller than $5$. Measuring $n_\text{t}$ and $t_\text{cg}$
independently gives $\tau=1.11\pm0.03$ and $\beta=2.88\pm0.02$, see
the insert in figure \ref{fig4-5}.

For the Monte Carlo algorithm, each sweep ideally contains
$(L/\Lambda)^2$ individual Monte Carlo updates. Each Monte Carlo
update consists of time stepping a sub lattice of size
$\Lambda\times\Lambda$. Hence, the cost of a Monte Carlo update is
$ab\Lambda^{\alpha_\text{TS}}$ when using equation
(\ref{timets}). However, each time stepping of a sub lattice is
followed by solving the Kirchhoff equations for the {\it entire\/}
lattice in order to determine $Q$ for the trial configuration. The
cost of this operation is $b L^\beta$. The time per Monte Carlo sweep
is then
\begin{equation}
\label{eq:sweept}
T_\text{MC}=\left(\frac{L}{\Lambda}\right)^2\ 
\left[4ab\Lambda^{\alpha_\text{TS}}
+bL^\beta\right]=4ab\Lambda^{\alpha_\text{TS}-2}L^2+\frac{b}{\Lambda^2}\ 
L^{2+\beta}\;,
\end{equation}
where $\alpha_\text{TS}=3.99$ and $\beta=2.88$. The factor ``4"
signifies that we time step the sub lattice for four pore volumes. By
setting $a\approx 10$ and $\Lambda=20$, the first term will dominate
compared to the second term on the right hand side of this equation if
$4a\Lambda^{\alpha_\text{TS}} \approx 6.4\times 10^6> L^{2.88}$ or
$L>230$ where the second term, which scales as $L^{4.88}$, 
starts dominating. It is this behavior
we see in figure \ref{fig4-5}: the computational time in the Monte
Carlo method scales according to the first term, i.e., as $L^2$.

Hence, we summarize: The time stepping procedure scales as $L^{3.99}$
whereas the Monte Carlo algorithm scales as $L^{1.98}$, as shown in
figure \ref{fig4-5}.

\subsection{Limitations}
\label{sub:limitations}

A closer inspection of figures \ref{fig4-1} to \ref{fig4-4} shows that
the match between the Monte Carlo and the time stepping procedures is
good but not perfect. In this section we discuss the discrepancies
between the two methods quantitatively.

\begin{figure}[h]
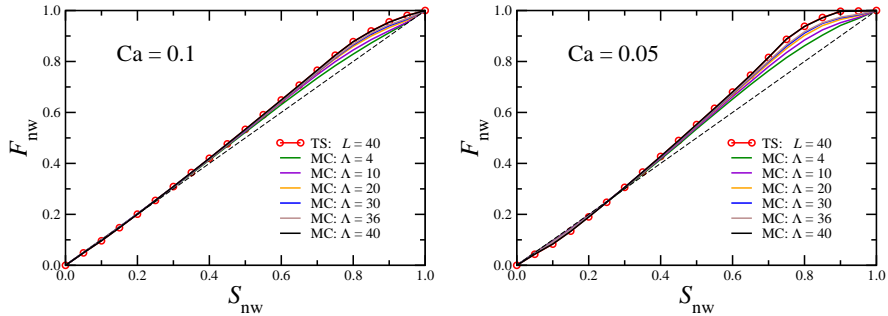

\centerline{\hfill
\includegraphics[width = 0.46\textwidth,clip]{fig4-6a.eps}\hfill
\includegraphics[width = 0.46\textwidth,clip]{fig4-6b.eps}\hfill}
\caption{\label{fig4-6} Non-wetting fractional flow $F_\text{nw}$ as a
  function of non-wetting saturation $S_\text{nw}$ for time stepping
  compared to Monte Carlo for different sub-network sizes ($\Lambda$)
  in the constant $Q$ ensemble. The size of the network, $L$, is $40$
  for both Monte Carlo and time stepping.}
\end{figure}

\begin{figure}[h]
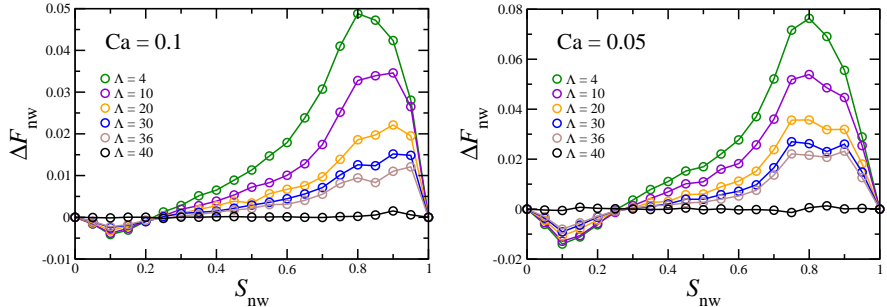

\centerline{\hfill
\includegraphics[width = 0.46\textwidth,clip]{fig4-7a.eps}\hfill
\includegraphics[width = 0.46\textwidth,clip]{fig4-7b.eps}\hfill}
\caption{\label{fig4-7} The difference of the non-wetting fractional
  flow ($\Delta F_\text{nw}$) between time stepping and Monte Carlo
  for different values of $\Lambda$ is plotted as a function of
  $S_\text{nw}$. $\Delta F_\text{nw}$ fluctuates around zero for
  $\Lambda = L$ and a systematic increase is observed with the
  decrease in $\Lambda$ for the whole range of $S_\text{nw}$.}
\end{figure}

We show in figure \ref{fig4-6} the non-wetting fractional flow for a
$40\times 40$ network using both time stepping and Monte Carlo with
sub network size $\Lambda$ ranging from $4$ to $40$. Notice that we
also consider the sub-network size $40$ which is equal to $L$. The
calculations here are done in the constant $Q$ ensemble with a
capillary number Ca equal to $0.1$ or $0.05$. As we see, there is a
systematic deviation between the time stepping and the Monte Carlo
results that increases with increasing non-wetting saturation
$S_\text{nw}$. This deviation is highlighted in figure \ref{fig4-7}
where the difference between the time stepping and the Monte Carlo
results for different $\Lambda$ is shown. We note that the difference
between the Monte Carlo and the time stepping decreases with
increasing capillary number Ca. This is, however, to be expected, as
for infinite $\text{Ca}$, any curve, Monte Carlo or time stepping,
must fall on the diagonal of figure \ref{fig4-6}.

\begin{figure}[h]
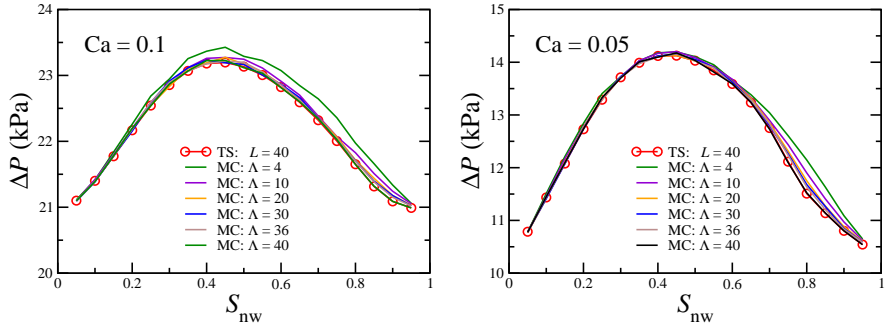

\centerline{\hfill
\includegraphics[width = 0.46\textwidth,clip]{fig4-8a.eps}\hfill
\includegraphics[width = 0.46\textwidth,clip]{fig4-8b.eps}\hfill}
\caption{\label{fig4-8} Pressure difference $\Delta P$ as a function
  of non-wetting saturation $S_\text{nw}$ for time stepping compared
  with Monte Carlo for different sub-network sizes ($\Lambda$) in the
  CQ ensemble. The size of the network, $L$, is $40$ for both Monte
  Carlo and time stepping.}
\end{figure}

In figure \ref{fig4-8} we show the discrepancy between the pressure
drop $\Delta P$ using time stepping and Monte Carlo for different sub
lattice size $\Lambda$. The systematics seen in the fractional flow
data, figures \ref{fig4-6} and \ref{fig4-7}, where the difference
grows with increasing non-wetting saturation is much less pronounced
in this case.

\begin{figure}[h]
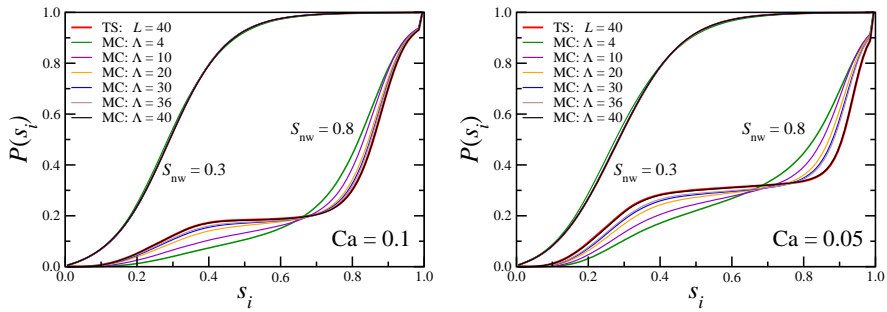

\centerline{\hfill
\includegraphics[width = 0.46\textwidth,clip]{fig4-9a.eps}\hfill
\includegraphics[width = 0.46\textwidth,clip]{fig4-9b.eps}\hfill}
\caption{\label{fig4-9} Comparison of cumulative distribution $P(s_i)$
  of link-saturation $s_i$ for time stepping and for Monte Carlo with
  different sub-system sizes. For $S_\text{nw}=0.3$, $P(s)$ for Monte
  Carlo match with time stepping for all the subsystem sizes, whereas
  for $S_\text{nw}=0.8$, a systematic difference in $P(S)$ is observed
  for $\Lambda<L$.}
\end{figure}

\begin{figure}[h]
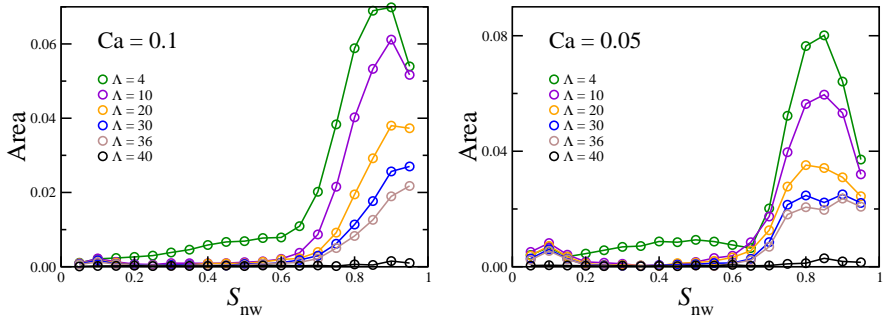

\centerline{\hfill
\includegraphics[width = 0.46\textwidth,clip]{fig4-10a.eps}\hfill
\includegraphics[width = 0.46\textwidth,clip]{fig4-10b.eps}\hfill}
\caption{\label{fig4-10} Area between the $P(s_i)$ curves
(Figure \ref{fig4-9}) for time stepping and for Monte Carlo 
with different sub-system sizes as a function of $S_\text{nw}$.}
\end{figure}

In figure \ref{fig4-9}, we show histograms over the non-wetting
saturation of the links. That is, we measure how much non-wetting
fluid each link contains. When the overall non-wetting saturation
$S_\text{nw}=0.3$, there is essentially no difference between the time
stepping and the Monte Carlo result. However, for $S_\text{nw}=0.8$,
there is a difference that depends on the sub lattice size
$\Lambda$. This difference, measured as the area between the time
stepping and the Monte Carlo histograms, is shown in figure
\ref{fig4-10} as a function of $S_\text{nw}$. The picture seen here
resembles that seen for the non-wetting fractional flow (figure
\ref{fig4-6}): the difference grows with increasing $S_\text{nw}$.

When the non-wetting saturation $S_\text{nw}$ is small, the
non-wetting fluid will form bubbles or small clusters surrounded by
the wetting fluid. As $S_\text{nw}$ is increased, these clusters grow
in size until there is a percolation-type transition where the
wetting fluid starts forming clusters surrounded by the non-wetting
fluid. This scenario has been studied experimentally by Tallakstad et
al.\ \cite{tkrlmtf09,tlkrfm09}. They argued that there is a length
scale $l^*$. Clusters that are larger than this length scale will
move, whereas clusters that are smaller will be held in place by the
capillary forces. The Monte Carlo algorithm calls for selecting a sub
network which is then ``lifted" out of the system, ``folded" into a
torus and then time stepped. The boundaries of the sub network will
cut through clusters and mobilize these. This changes the cluster
structure from that of the time stepping procedure.

\begin{figure}[h]
\centerline{\hfill
\includegraphics[width = 0.24\textwidth,clip]{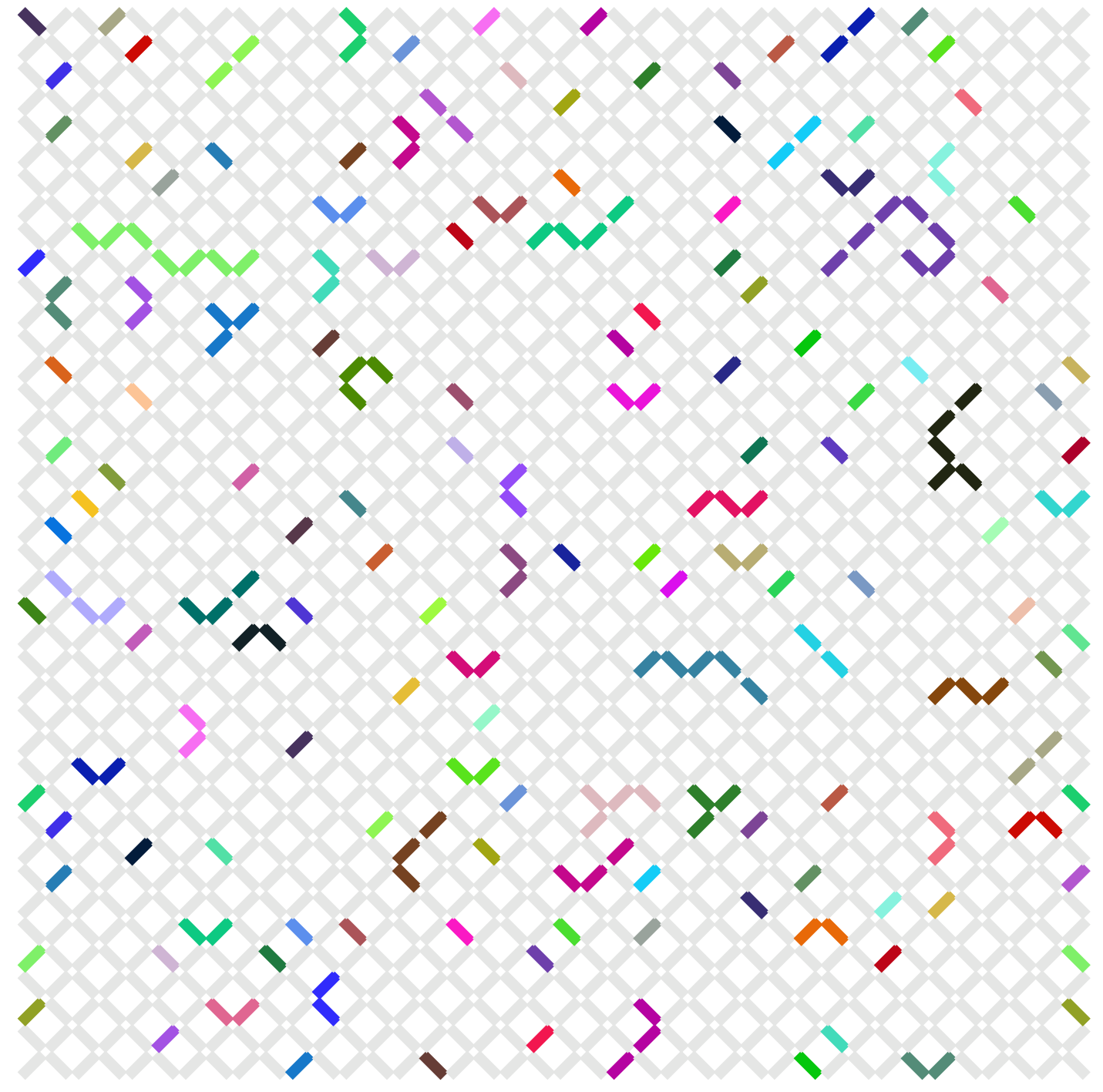}\hfill
\includegraphics[width = 0.24\textwidth,clip]{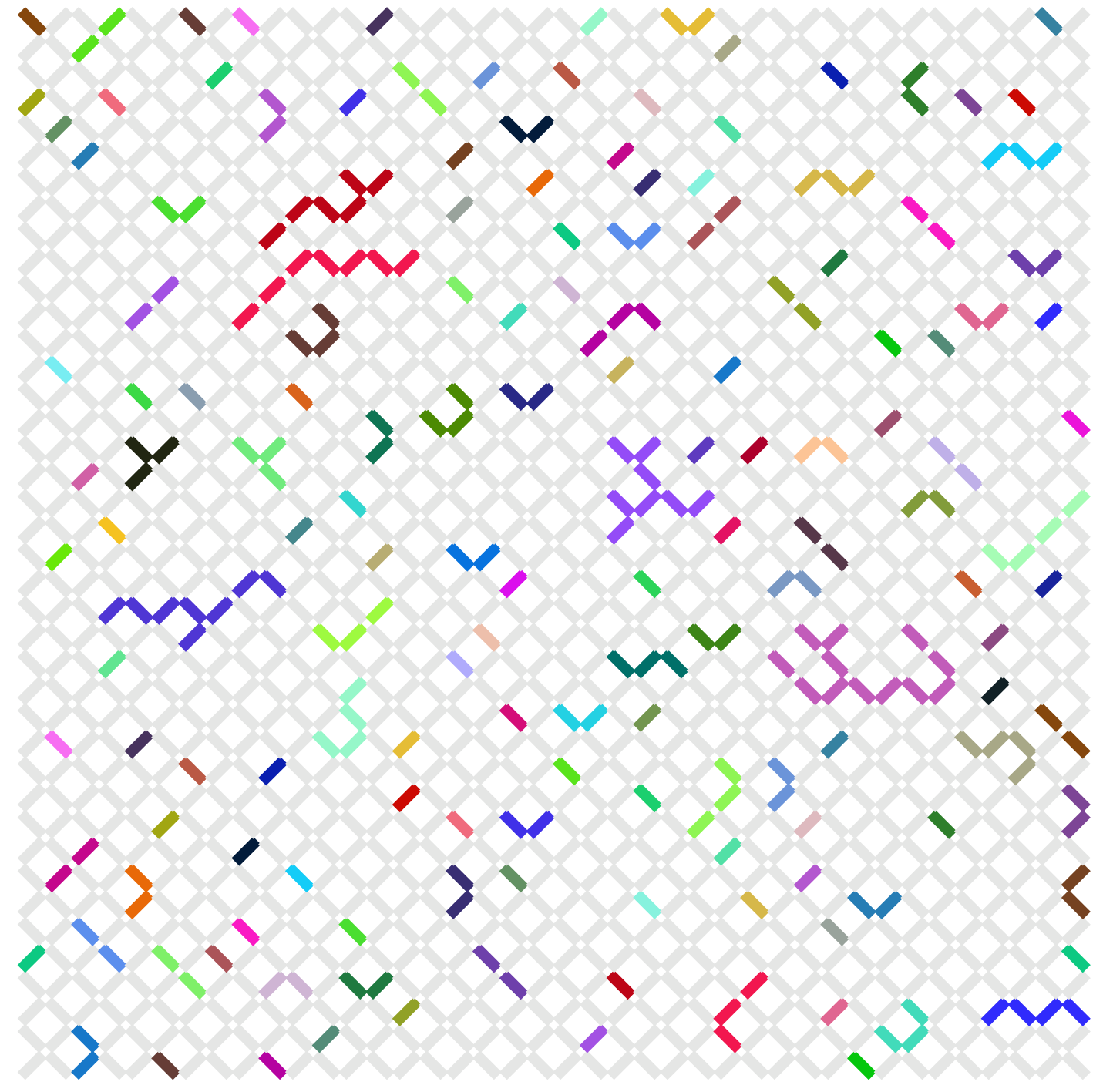}\hfill
\includegraphics[width = 0.24\textwidth,clip]{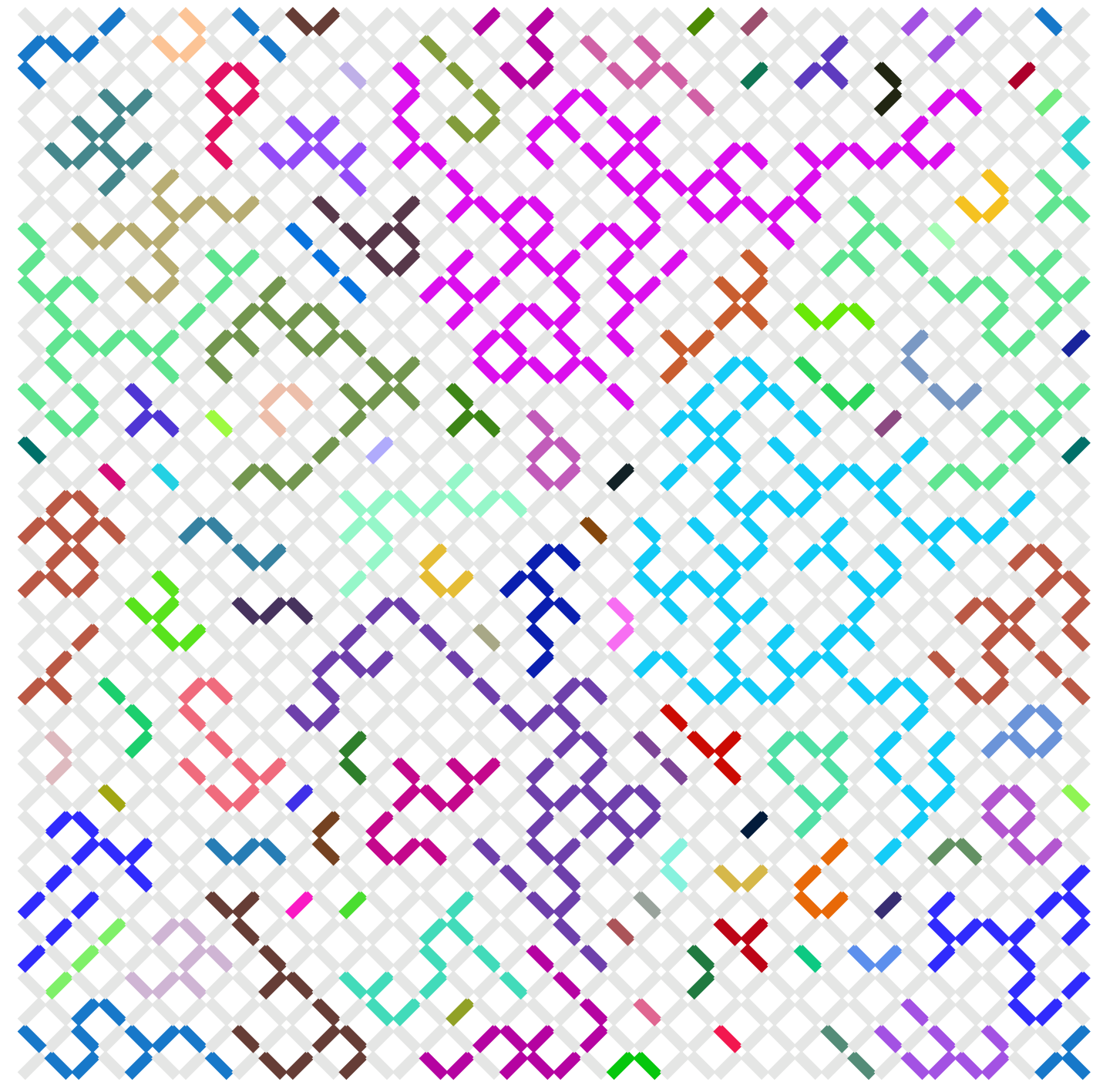}\hfill
\includegraphics[width = 0.24\textwidth,clip]{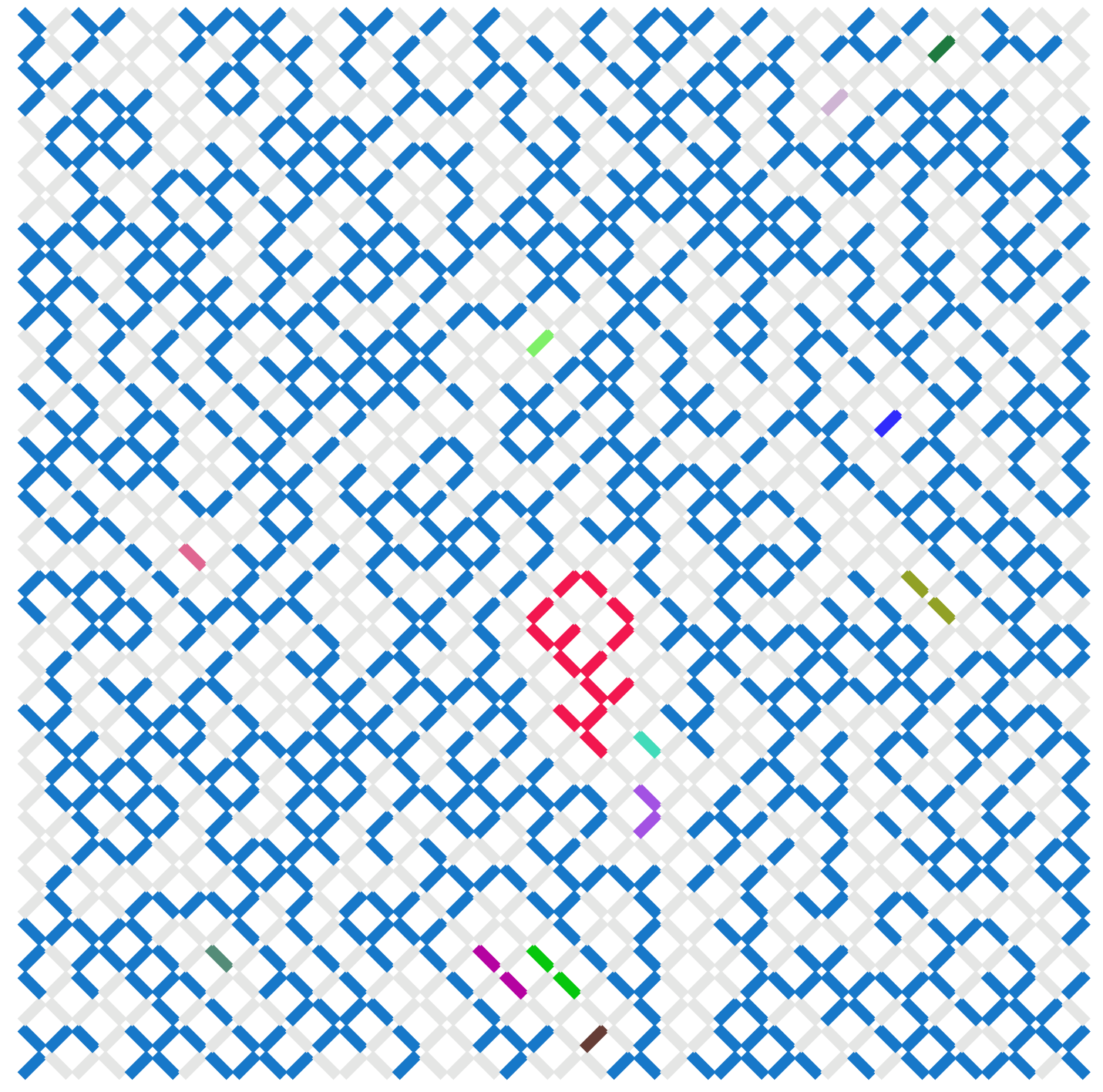}\hfill}
\centerline{\hfill MC, $S_\text{nw}=0.7$ \hfill \hfill TS, $S_\text{nw}=0.7$ \hfill \hfill MC, $S_\text{nw}=0.8$ \hfill \hfill TS, $S_\text{nw}=0.8$ \hfill}
\caption{\label{fig4-11} Typical non-wetting clusters for Monte Carlo
  (MC) and time stepping (TS) at $\text{Ca}=0.05$. The network is of
  $40\times 40$ links and the sub network size for Monte Carlo is
  $20\times 20$ links. Each cluster is marked with different colours so
  that the structure is readily visible.}
\end{figure}

In order to investigate this we have studied the cluster structure in
the model under Monte Carlo and time stepping. In order to do this, we
identify the non-wetting clusters. To do this, two nodes are
considered to be part of the same cluster if the link between them has
a non-wetting saturation more than a threshold value, a clip-threshold
$c_t$. Here we use a clip threshold equal to $c_t = 0.9$
\cite{rh06}. In figure \ref{fig4-11}, we show typical cluster
structures for two different non-wetting saturations obtained with
Monte Carlo and with time stepping. For $S_\text{nw}=0.7$, the
non-wetting clusters are still quite small and there is no discernable
difference between the configurations obtained with time stepping and
with Monte Carlo. However, for $S_\text{nw}=0.8$, there is one
dominating cluster in the time stepping case whereas the clusters are
more broken up in the Monte Carlo case.

\begin{figure}[h]
\includegraphics[width = 0.66\textwidth,clip]{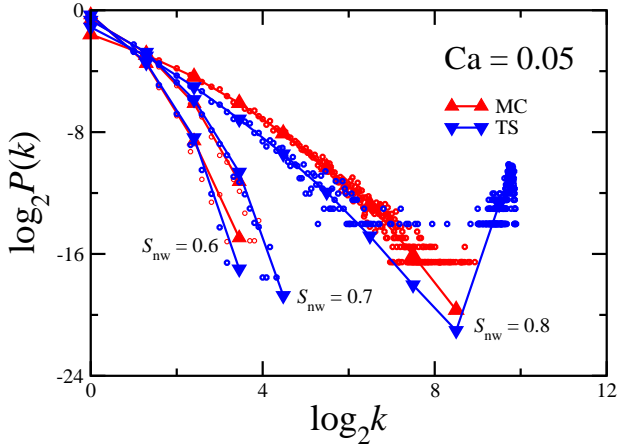}
\caption{\label{fig4-12} Cluster size distribution
  $P(k)=N_k/N_{total}$ versus cluster size $k$ for time stepping and
  Monte Carlo. The blue circles signify the Monte Carlo data and the
  red circles the time stepping data. The red and blue curves with
  triangles pointing upwards or downwards signify the Monte Carlo and
  time stepping data after logarithmic binning. Here $L=40$ and
  $\Lambda=20$. The data are averaged over 10 samples.}
\end{figure}

We measure this qualitative difference in cluster structure for
$S_\text{nw}=0.8$ by recording the cluster size distribution for the
two types of updating, see figure \ref{fig4-12}. When following the
time stepping procedure, we run the system for $500$ pore
volumes. During the last $125$ pore volumes injected ($1/4$th of the
total), we measure the cluster size distribution after passing each
pore volume of fluids. When using Monte Carlo, we run the system for
400 Monte Carlo updates. We record the cluster size distribution for
every of the last $100$ updates. In both the time stepping and Monte
Carlo runs, we average over $10$ samples. The number of links belong
to a cluster defines the size of that cluster. The total number of
clusters is $N_{total}$ and the number of clusters of size $k$ that we
record is $N_k$. We show $P(k)=N_k/N_{total}$ in the figure. For
$S_\text{nw}=0.6$ and 0.7, there is no discernable difference in the
cluster structure between the Monte Carlo and the time stepping
procedures. However, for $S_\text{nw}=0.8$, there are differences. For
every $k$ the number of clusters during the Monte Carlo updating
procedure is larger than for the time stepping procedure, except for
the largest clusters, the percolating cluster seen in figure
\ref{fig4-11}. This supports the supposition that the Monte Carlo
breaks up the large non-wetting clusters.

Clearly, for the Monte Carlo algorithm to be perfected, this tendency of
chopping up large non-wetting clusters needs to be counteracted. Presumably,
this is a problem that decreases with increasing system and sub lattice size
as it is a boundary effect.  

\section{Conclusion}
\label{sec:conc}

We have in this work presented a new Monte Carlo algorithm for
immiscible two-phase flow in porous media under steady-state
conditions using network models. It is based on the Metropolis
transition probability (\ref{eq:metq}) which in turn is build upon the
configuration probability (\ref{eq:piintq}) which we derive here. By
steady-state conditions, we mean that the macroscopic parameters that
describe the flow such as pressure difference, flow rate, fractional
flow rate and saturation all have well defined means that stay
constant. On the pore level, however, clusters flow, merge, break up,
and so on. The flow may be anything but stationary. We described the
algorithm in Section \ref{subsub:implementation}.

Computationally, the Monte Carlo algorithm is very fast compared to
time stepping. We find that the time stepping procedure when
implemented on a square lattice demands a computing time that scales
as the linear size of the lattice, $L$, to the fourth power, whereas
the Monte Carlo method scales as the linear size to the second power,
see Section \ref{sub:cost}. However, there is another term that
contributes to the computing time in the Monte Carlo procedure which
scales as $L^{4.88}$. This term has a prefactor associated with it
which is very small compared to the other term scaling as $L^2$. For
$L$ up to about 230, this term is small compared to the first one.

\subsection{Open Questions}
\label{sub:open}

There are open questions with respect to the Metropolis Monte Carlo
approach that we present here. The most important step in the
direction of constructing such an approach is to identify the
configuration probability (\ref{eq:piintq}). The second most
important step is to provide a way to generate trial configurations
that obey the symmetry requirement (\ref{eq:sym}). Section
\ref{subsub:implementation} is concerned with this.

We see three challenges that will need to be overcome before the Monte
Carlo algorithm that we propose here is fully capable of replacing
time stepping.

\begin{itemize}
\item The Monte Carlo algorithm needs to be generalized to irregular
  networks, e.g., those based on reconstructed porous media
  \cite{bbdgimpp13}.\\
\item The necessity to solve the Kirchhoff equations for the entire
  pore network once for every Monte Carlo update will slow down the
  algorithm when it is implemented for large systems. Ideally, one
  should find a way to circumvent this necessity.\\
\item The Monte Carlo algorithm has a tendency to break up large
  non-wetting clusters as described in Section \ref{sub:limitations}.
  This is a problem for large non-wetting saturations. It is most
  probably a boundary effect that comes from the way the sub networks
  are constructed. However, it needs to be overcome if the algorithm
  is to be useful for the entire range of saturations.\\
\end{itemize}

Overcoming these three challenges will allow network models to take 
advantage to the full of the ongoing revolution in pore space
characterization.

We have in this article presented a first attempt at constructing a 
Markov Chain Monte Carlo algorithm based on the configurational probability 
(\ref{eq:piintq}).  There is no reason not to believe that other ways of 
constructing such Monte Carlo algorithms might be possible that are both 
faster and do not pose the challenges listed above.  

\acknowledgement{IS thanks VISTA, a collaboration between Statoil and
  the Norwegian Academy of Science and Letters for financial support.
  SS and AH thank the Beijing Computational Science Research Center
  and its director, Professor Hai-Qing Lin, for financial support and
  for providing an excellent atmosphere for doing science.  We thank
  Eirik Grude Flekk{\o}y, Knut J{\o}rgen M{\aa}l{\o}y, Miguel Rubi and
  Marios Valavanides for many interesting discussions.}


\end{document}